\documentclass[11pt]{article}
\usepackage[a4paper,margin=24mm]{geometry}
\usepackage{amsmath,amssymb,graphicx,booktabs,array}
\usepackage[font=small,labelfont=bf]{caption}
\usepackage[numbers,sort&compress]{natbib}
\usepackage[colorlinks=true,linkcolor=blue,citecolor=blue,urlcolor=blue]{hyperref}
\hypersetup{pdftitle={LoRA Adaptation Strength in Aurora-WRF: Trade-offs in Regional Heavy-Rainfall Forecasting},pdfauthor={Boyan Liu and Xiaoyuan Zhang}}
\usepackage{microtype}
\makeatletter
\let\tableinput\@@input
\makeatother
\graphicspath{{figures/}}
\title{LoRA Adaptation Strength in Aurora--WRF:\\Trade-offs in Regional Heavy-Rainfall Forecasting}
\author{Boyan Liu \quad Xiaoyuan Zhang\raisebox{.6ex}{\normalsize *}\\[4pt]
\normalsize Zhongguancun Academy, Beijing, China\\[2pt]
\small \href{mailto:liberty151208@163.com}{\texttt{liberty151208@163.com}}\quad
\href{mailto:zhangxiaoyuan@bza.edu.cn}{\texttt{zhangxiaoyuan@bza.edu.cn}}\\[2pt]
\small * Corresponding author}
\date{September 2026}
\begin{document}
\maketitle
\begin{abstract}
We investigate how parameter-efficient adaptation of an atmospheric foundation model affects downstream regional precipitation in a controlled Aurora--WRF coupling experiment over the Beijing--Tianjin--Hebei region. A single-step, precipitation-weighted LoRA adaptation of AuroraPretrained is trained on May--September 2020--2021 and selected using 2022 validation data. Five inference-time adaptation strengths are coupled to an unchanged 9-km WRF configuration and compared with a GFS-driven control under common ERA5 initialization and auxiliary-field rules. Evaluation uses 17 initializations in 2023, hourly ERA5 precipitation, and conservative remapping to a fixed 356-cell, 0.25-degree verification grid. Heavy rainfall is defined exclusively as at least 50 mm in a continuous 24-hour window, advanced hourly. For window starts from 0 to 48 hours, unadapted Aurora attains the highest mean critical success index (CSI), 0.216, compared with 0.144 for the GFS control. An intermediate LoRA strength of 0.75 reduces the Aurora-driven 24-hour precipitation mean squared error by 30.7\% and brings pooled frequency Bias from 1.510 to 1.024, but lowers CSI to 0.182. GFS retains the lowest whole-domain error. These results characterize a trade-off between rainfall detection, frequency calibration, and intensity error, showing the importance of evaluating adaptation strength against multiple downstream criteria within a common coupling framework.
\end{abstract}

\section{Introduction}
Atmospheric foundation models offer a computationally efficient source of large-scale forecast states. Aurora supports adaptation across atmospheric prediction tasks using a shared pretrained representation \citep{aurora}. Regional numerical models such as the Weather Research and Forecasting model (WRF) resolve additional dynamics and parameterize processes that are absent from a coarse atmospheric state forecast \citep{wrf}. Coupling the two therefore offers a practical route to regional precipitation prediction without requiring the AI model to predict precipitation directly.

The coupling itself is not new. For example, Pangu-Weather-driven WRF has been studied for extreme precipitation in North China \citep{panguwrf}. The question considered here is narrower: how does the magnitude of a learned, low-rank atmospheric-state adjustment affect rainfall after integration by a fixed regional numerical model? Low-rank adaptation (LoRA) is a parameter-efficient training mechanism \citep{lora}, but the training target and downstream verification target are different in this application. A reduction in normalized atmospheric-state error is not equivalent to a reduction in rainfall error, nor does it guarantee improved threshold-event detection.

We study a single trained LoRA checkpoint at five strengths, including the unadapted and fully adapted endpoints. Our contributions are: (i) a reproducible, controlled boundary-forcing comparison with explicit auxiliary-field provenance; (ii) a fixed verification protocol separating hourly precipitation errors from hourly moving 24-hour heavy-rainfall events; and (iii) a multi-metric analysis showing how aggregate improvements depend on adaptation strength and case composition. All reported predictions use the same single-step adaptation experiment. This paper does not report multi-step training, spectral objectives, ensemble forecasting, or a new WRF dynamical core.

\section{Data and experimental design}
\subsection{Training, validation, and forecast cases}
The local adaptation data comprise ERA5 fields for May--September of 2020--2023. ERA5 is a model-based reanalysis, not direct gauge truth \citep{era5}. Training uses 1,224 single-step samples from 2020--2021; validation uses 612 samples from 2022. A sample supplies two atmospheric states six hours apart and predicts the subsequent six-hour state. The data region spans 33--45$^\circ$N and 108--123.75$^\circ$E on a 0.25$^\circ$ grid. The regional Aurora implementation produces 48 by 64 spatial points after its latitude crop.

Downstream rainfall evaluation uses 17 initializations at 00 UTC in 2023 (Appendix~\ref{app:cases}). The sample includes wet and dry conditions and is event-enriched: five dates were selected by a reference-only check for at least one cell with 24-hour precipitation reaching 50 mm. It is not a random sample of forecast dates. The 96-hour forecast periods overlap for 7 and 10 August (24 hours) and for 19 and 21 August (48 hours); the initializations are therefore not treated as 17 independent meteorological events.

The local LoRA training and checkpoint selection do not use 2023. Nevertheless, these forecast cases have been inspected repeatedly during development, and adaptation strengths are compared retrospectively. They are diagnostic cases, not an untouched final test. The local year split also does not by itself establish the complete training-data history of the upstream foundation checkpoint.

\subsection{Forecast configurations}
Six drivers are compared: GFS and Aurora at LoRA strengths $\alpha\in\{0,0.25,0.50,0.75,1\}$. Each driver supplies future large-scale atmospheric states to the same WRF configuration. The evaluation comprises 102 forecast archives with hourly precipitation through 96 hours; hours 72--96 are used only for the extended window diagnostic.

The principal analysis uses window starts $s=0,\ldots,48$ hours and thus precipitation ending no later than 72 hours. A supplementary analysis retains $s=0,\ldots,72$, requiring forecasts through 96 hours. Both are reported, and the latter is not relabeled as a forecast ending at 72 hours.

\section{Methods}
\subsection{Precipitation-weighted, single-step adaptation}
The model is initialized from the official AuroraPretrained checkpoint and fresh LoRA parameters, not from an earlier small-sample adapted checkpoint. All pretrained parameters remain frozen. The implementation trains 2,850,816 LoRA parameters with four-GPU distributed data parallelism, a global batch size of four, AdamW, learning rate $5\times10^{-5}$, weight decay $10^{-4}$, gradient clipping at one, seed 42, and five epochs. The selected checkpoint minimizes its own precipitation-weighted 2022 validation state loss, including the epoch-zero reference. It does not minimize a 2023 rainfall score.

Let $e_j$ denote the mean normalized absolute state error at spatial point $j$: the four surface-variable error maps and the five pressure-level-averaged atmospheric-variable error maps each receive equal weight. The pixel weights and loss are
\begin{align}
 u_j&=\min\{10,\,1+0.5\log(1+\frac{P^{\rm ref}_{6,j}}{1\,\mathrm{mm}})
       +4\,\mathbf{1}[P^{\rm ref}_{24,j}\geq50\,\mathrm{mm}]\},\\
 w_j&=u_j/\langle u\rangle_j,\qquad
 \mathcal L_{\rm state}=\langle w_j e_j\rangle_j .
\end{align}
Here $P^{\rm ref}_{6}$ is the reference accumulation over the predicted six-hour interval, and $P^{\rm ref}_{24}$ is the reference accumulation over the 24 hours ending at the target state time. These quantities weight the training loss; they are not model precipitation outputs or future inputs at inference. The selected AuroraPretrained interface predicts four surface variables and five atmospheric variables on 13 pressure levels. It has no precipitation output head in this experiment.

For a linear-layer weight, inference-time strength acts as
\begin{equation}
 W(\alpha)=W_0+\alpha s_{\rm L}BA,
\end{equation}
where $W_0$ is frozen, $A$ and $B$ are the trained low-rank factors, and $s_{\rm L}$ is the module's internal scale. Thus $\alpha=0$ disables the learned increment and $\alpha=1$ uses its full value. Intermediate strengths do not mix two completed forecasts. Nonlinear and autoregressive model operations mean that the resulting atmospheric fields, and subsequently WRF rainfall, need not vary monotonically with $\alpha$.

The model is trained for one six-hour prediction and then run autoregressively for 12 or 16 steps to reach 72 or 96 hours. WRF is not differentiable within this training pipeline: no precipitation loss is backpropagated through WRF. The implementation uses Microsoft Aurora package version 2.0.1.

\subsection{Common initialization and boundary construction}
All six drivers share ERA5 initialization at $t_0-6$ hours and ERA5 atmospheric forcing through $t_0$, providing a six-hour adjustment period. For times after $t_0$, the principal atmospheric fields come from the corresponding GFS or Aurora forecast at six-hour intervals. WRF consumes the time-dependent lateral boundary conditions prepared from those fields; it does not require knowledge of future ERA5 atmospheric states.

To match the Aurora interface, the GFS control is restricted to four surface fields and five variables at 13 pressure levels, rather than using its complete operational variable set. The surface fields are 2-m temperature, 10-m zonal and meridional wind, and mean sea-level pressure; upper-air fields are temperature, zonal and meridional wind, specific humidity, and geopotential. The pressure levels are 50, 100, 150, 200, 250, 300, 400, 500, 600, 700, 850, 925, and 1000 hPa. This is a controlled large-scale forcing comparison, not a full operational GFS-to-WRF baseline.

The coupling clips negative forecast specific humidity to zero. Its southern padding row repeats the last Aurora output row outside the verification domain. Physical units and checkpoint hashes are documented in the ancillary reproducibility notes; the WRF configuration is specified below.

\begin{table}[t]
\centering\small
\caption{Shared auxiliary-field rules. Initialization fields are not future observations.}
\label{tab:aux}
\begin{tabular}{>{\raggedright\arraybackslash}p{.25\linewidth}>{\raggedright\arraybackslash}p{.66\linewidth}}
\toprule
Field & Source and treatment\\\midrule
Soil temperature/moisture, skin temperature, SST, sea ice, snow & ERA5 at the start of the adjustment period; prescribed auxiliary fields persist. The land-surface model evolves its internal state; SST updating is disabled.\\
Surface pressure $p_s$ & After $t_0$, $\widehat p_s(t)=[p_s(t_0)/p_{\rm msl}(t_0)]\widehat p_{\rm msl}(t)$.\\
2-m specific humidity $q_2$ & Derived from ERA5 dew point and pressure at $t_0$; scaled subsequently by forecast lowest-level specific humidity relative to its initial value and capped at saturation.\\
Cloud liquid and ice & ERA5 cloud fields only at initialization; subsequent prescribed cloud boundary mixing ratios are zero.\\
Static fields & Common land--sea mask, orography, soil categories, and WPS geography.\\\bottomrule
\end{tabular}
\end{table}

Table~\ref{tab:aux} summarizes fields absent from the selected Aurora interface. Auditing the forcing code found no direct injection of post-$t_0$ ERA5 atmospheric states into the forecast boundary fields. This does not make the setup operational: ERA5 initialization is retrospective, and the auxiliary approximations may affect downstream precipitation. Sharing those rules removes one source of asymmetry but does not prove equivalence to each model's native operational coupling.

\subsection{WRF configuration and spatial matching}
The regional model is WRF-ARW 4.8.0 on a single 9-km Lambert conformal domain, with 100 by 100 mass-grid cells, 45 vertical levels, a 50-hPa top, and a 45-s time step. The projection reference is 39.5$^\circ$N, 115.875$^\circ$E, with standard parallels at 30$^\circ$ and 60$^\circ$N. The fixed configuration uses \texttt{cu\_physics=3}, the Grell--Freitas--Li implementation in this WRF version, whose scheme family is described by \citet{gf,gfl}. Other options are Thompson microphysics (8), YSU boundary layer (1), revised MM5 surface layer (1), Noah land surface (2), and RRTMG shortwave/longwave radiation (4). No urban scheme is enabled. These are configuration and coupling choices, not modifications to the WRF dynamical equations.

Hourly precipitation is obtained from \texttt{RAINC+RAINNC}, subtracting its value at $t_0$ to exclude adjustment-period rain. The model is not a 1-km forecast system in this study. Verification uses a fixed 356-cell ERA5 0.25$^\circ$ mask. Native WRF rainfall is conservatively remapped by source/target grid-polygon overlap in a Lambert azimuthal equal-area projection. The outer five WRF cells are excluded when constructing source polygons; accepted target cells have at least 99.9\% coverage within the common geographic mask established in the earlier configuration comparison. The remapping is area-overlap averaging, not bilinear interpolation or an increase in effective resolution.

If $a_{jg}$ is the overlap area between target cell $j$ and WRF cell $g$, the mapped accumulation is $\widehat C_j=\sum_g a_{jg}C_g/a_j$, where $a_j$ is target-cell area. Spatial averaging can smooth extremes and alter threshold exceedances, so performance here applies to the verification support, not to native 9-km pointwise skill or unresolved 1-km structure.

\subsection{Verification definitions}
Let $C_{d,i,j}(h)$ be cumulative forecast precipitation from initialization to lead $h$, for driver $d$, initialization $i$, and target cell $j$. For integer window start $s$,
\begin{equation}
 P^{24}_{d,i,j}(s)=C_{d,i,j}(s+24)-C_{d,i,j}(s)
                 =\sum_{k=s+1}^{s+24}P^1_{d,i,j}(k).
\end{equation}
Reference accumulations use the same hours. A heavy-rainfall event is $P^{24}\geq50$ mm. The window advances by one hour; the accumulation duration remains 24 hours. In particular, no threshold of $50/24$ mm is applied to hourly precipitation and no hourly value is multiplied by 24.

For each driver and window, hits $H_{d,s}$, false alarms $F_{d,s}$, and misses $M_{d,s}$ are first summed across all initializations and the 356 cells. The headline scores are
\begin{align}
 \mathrm{CSI}_{d,s}&=\frac{H_{d,s}}{H_{d,s}+F_{d,s}+M_{d,s}},\\
 \overline{\mathrm{CSI}}_d&=\frac{1}{|\mathcal V_d|}\sum_{s\in\mathcal V_d}\mathrm{CSI}_{d,s},\\
 B_d&=\frac{\sum_s(H_{d,s}+F_{d,s})}{\sum_s(H_{d,s}+M_{d,s})},
\end{align}
where $\mathcal V_d$ contains that driver's defined windows. This CSI is not the mean of initialization-level CSI. Frequency Bias is pooled across all windows, not an arithmetic mean of window ratios. Pooled probability of detection (POD) is $\sum H/\sum(H+M)$, and pooled false-alarm ratio (FAR) is $\sum F/\sum(H+F)$. Zero denominators remain undefined, not zero or one. False alarms in windows without reference events remain in the pooled counts.

Whole-domain continuous errors use normalized target-cell areas $\omega_j$:
\begin{equation}
 \mathrm{MSE}_{d,s}^{24}=\frac{1}{N}\sum_{i=1}^N\sum_j\omega_j
 [P^{24}_{d,i,j}(s)-P^{24}_{{\rm ref},i,j}(s)]^2,
 \qquad \sum_j\omega_j=1.
\end{equation}
MAE substitutes absolute error. Classification counts are unweighted cell counts, consistent across drivers; continuous errors are area weighted. All valid cells contribute to continuous errors, not only reference heavy-rainfall cells. Actual hourly MAE/MSE are computed separately for leads 1--72 hours. Their units are also mm and mm$^2$, but they measure different accumulations from the 24-hour scores.

\subsection{Dispersion, case composition, and reproducibility}
For descriptive dispersion, we also calculate each initialization's own mean CSI over its valid windows, pooled Bias over its windows, and mean continuous errors. Standard deviations use the sample definition, denominator $n-1$, over defined initialization-level values. The effective $n$ is reported. Boxplots show medians, quartiles, 1.5-interquartile-range whiskers, and individual cases. These supplementary case means do not replace pooled headline scores, and their standard deviations are not confidence intervals or standard errors for the pooled estimator.

Adjacent 24-hour windows overlap, spatial cells are correlated, and some forecast periods overlap. No inference treating grid-cell/windows as independent observations is made. Leave-one-initialization-out diagnostics are retained in the reproducibility package as descriptive checks of sample sensitivity, not significance tests or a procedure for selecting a universally optimal strength.

For the lead-dependent pooled CSI and Bias, we estimate bootstrap standard deviations from 10,000 paired resamples of 15 groups, joining initializations whose 96-hour forecast periods overlap. Each resample draws 15 groups with replacement and retains all member initializations, cells, leads, and drivers together (seed 20260924). Scores are recomputed from the resampled contingency counts. Standard deviations use defined replicates only (at least 6422 per score and lead), so they are conditional when denominators vanish. These are descriptive bootstrap standard-error estimates, not between-case standard deviations or confidence intervals; grouping by overlap does not establish meteorological independence. MSE shading instead uses the sample standard deviation across the 17 initialization-level errors at each lead.

Archive identifiers and hashes for all 102 forecasts and their reference data, model identity, and scoring definitions are retained in the ancillary material. The analysis in this paper recomputes metrics from exported precipitation arrays, verifies the exported-array hash, and reproduces the previously archived 17-case H/F/M, mean CSI, Bias, MAE, and MSE. Scientific data, derived tables, and plotting code are separated from the manuscript sources.

\section{Results}
\subsection{Detection skill and precipitation error favor different strengths}
\begin{figure}[t]\centering
\includegraphics[width=\linewidth]{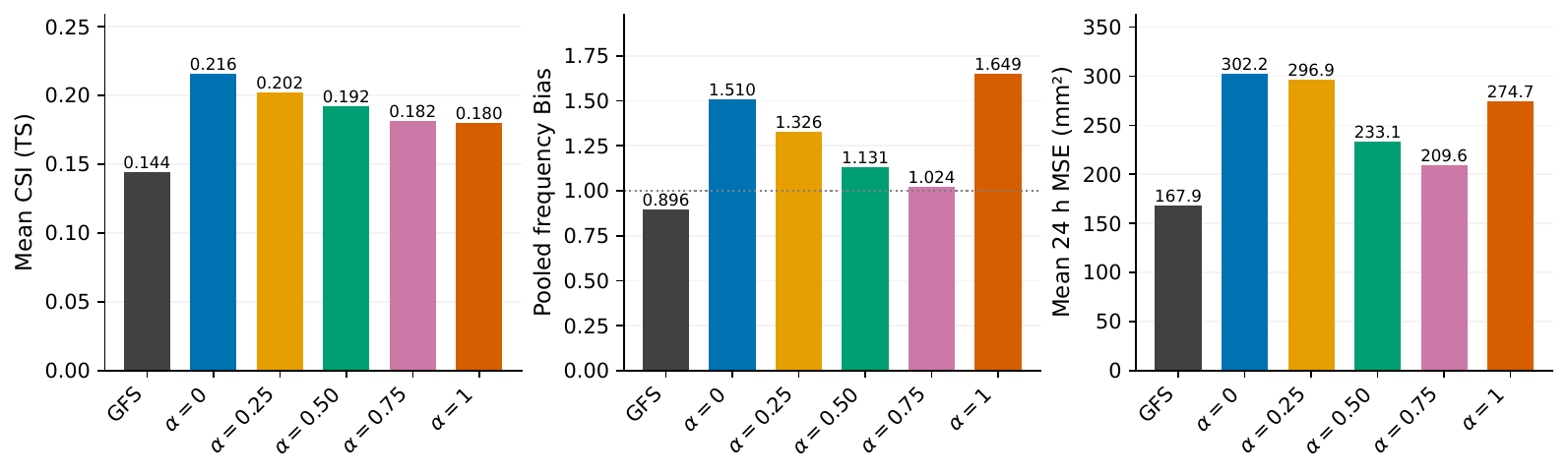}
\caption{Pooled main-scope performance for all six drivers. Bars deliberately show the established aggregate estimators without attaching case-level standard deviations to a different estimator. Case dispersion is reported separately in Table~\ref{tab:sd} and Figure~\ref{fig:boxes}.}
\label{fig:aggregate}\end{figure}

Table~\ref{tab:main} and Figure~\ref{fig:aggregate} show the main 0--48-hour window-start results. All drivers have 49 defined pooled CSI windows. Unadapted Aurora has the largest mean CSI (0.216) and POD (0.405), exceeding the controlled GFS values of 0.144 and 0.218. This improvement is accompanied by more predicted heavy-rainfall cell-windows: Bias is 1.510 instead of 0.896, and whole-domain 24-hour MSE is 302.20 instead of 167.88 mm$^2$.

\begin{table}[t]
\centering\small
\caption{Main 17-initialization results, 24-hour window starts 0--48 h. Each CSI averages 49 pooled window scores. MAE and MSE use all verification cells.}
\label{tab:main}
\begin{tabular}{lrrrrrr}\toprule
Driver & CSI & Bias & POD & FAR & MAE (mm) & MSE (mm$^2$)\\\midrule
\tableinput{tables/main.tex}
\bottomrule
\end{tabular}
\end{table}

Strength 0.75 provides a useful but conditional compromise. Its MSE is 209.57 mm$^2$, a 30.7\% reduction relative to unadapted Aurora, and pooled Bias is 1.024. However, CSI falls to 0.182 and POD to 0.286. The GFS control still has a smaller continuous error. Strength 0.50 has the smallest Aurora-driven MAE, 6.40 mm, and the smallest Aurora-driven hourly MSE, 1.53 mm$^2$; these need not coincide with the minimum 24-hour MSE. Full adaptation is not best: its Bias is 1.649 and FAR is 0.802, despite lower 24-hour MSE than unadapted Aurora. There is no single strength that dominates detection, frequency calibration, and intensity error.

\begin{figure}[t]\centering
\includegraphics[width=\linewidth]{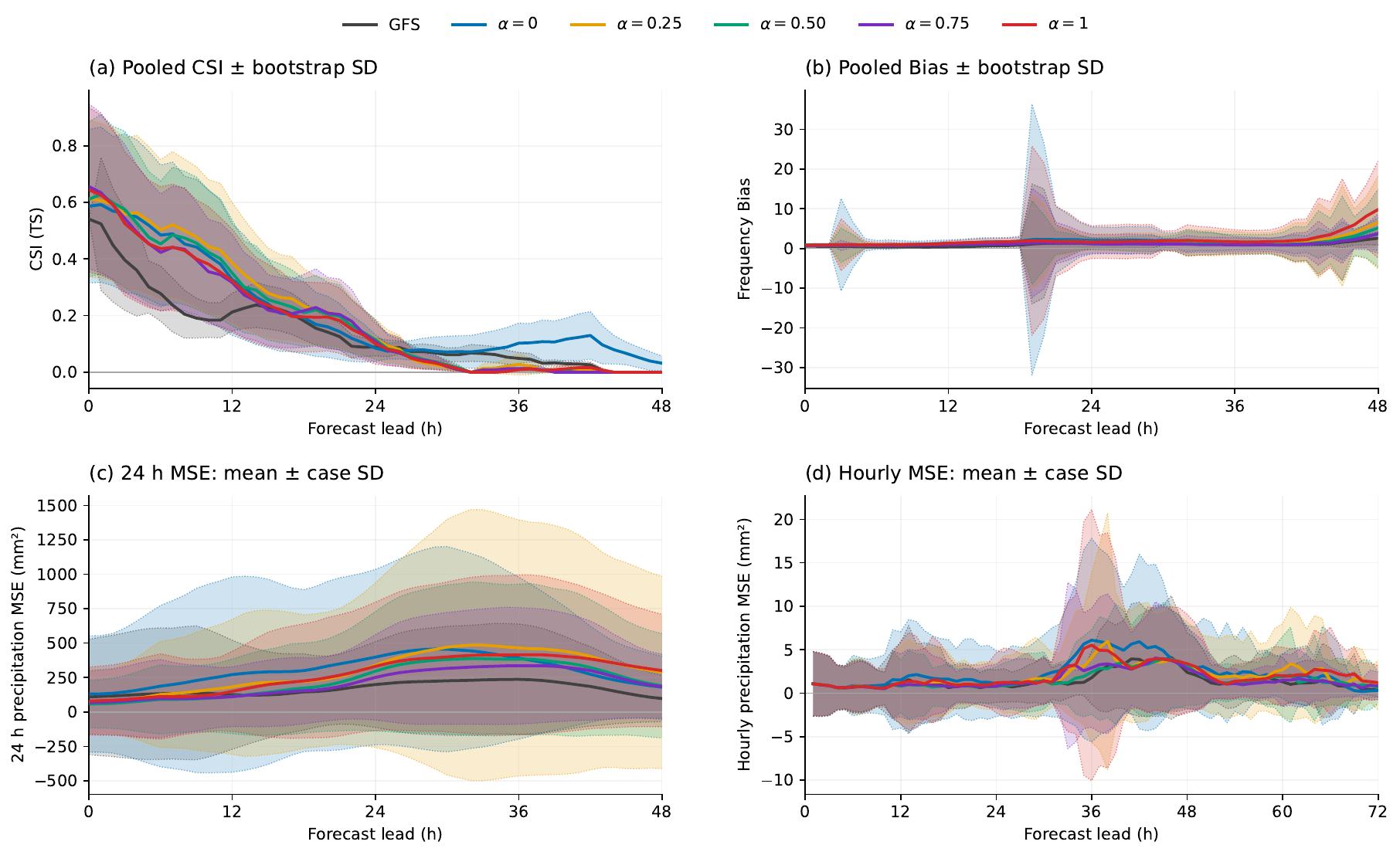}
\caption{Skill and errors versus forecast lead. Solid curves retain the pooled CSI/Bias and mean MSE estimators. Shading and its dotted boundaries show $\pm1$ bootstrap SD for CSI/Bias and $\pm1$ case SD for MSE; these are not confidence bands. Negative lower bounds reflect the symmetric SD construction, not negative scores. For the first three panels, lead is the 24-hour \emph{window start}: 0 denotes 0--24 h and 48 denotes 48--72 h. Hourly MSE uses interval ends 1--72 h, with no fabricated hour-zero error.}
\label{fig:curves}\end{figure}

The full hourly sequences in Figure~\ref{fig:curves} clarify that the bar-chart statistics are temporal summaries, not constant skill over the forecast. Moving-window detection generally weakens at later starts, while the continuous errors exhibit event-dependent peaks. A Bias near one should not be read as correct rain placement: substantial false alarms and misses can offset in a frequency ratio. None of the six mean CSI values reaches 0.25 in this main comparison.

\subsection{Initialization-level distributions}
\begin{figure}[t]\centering
\includegraphics[width=\linewidth]{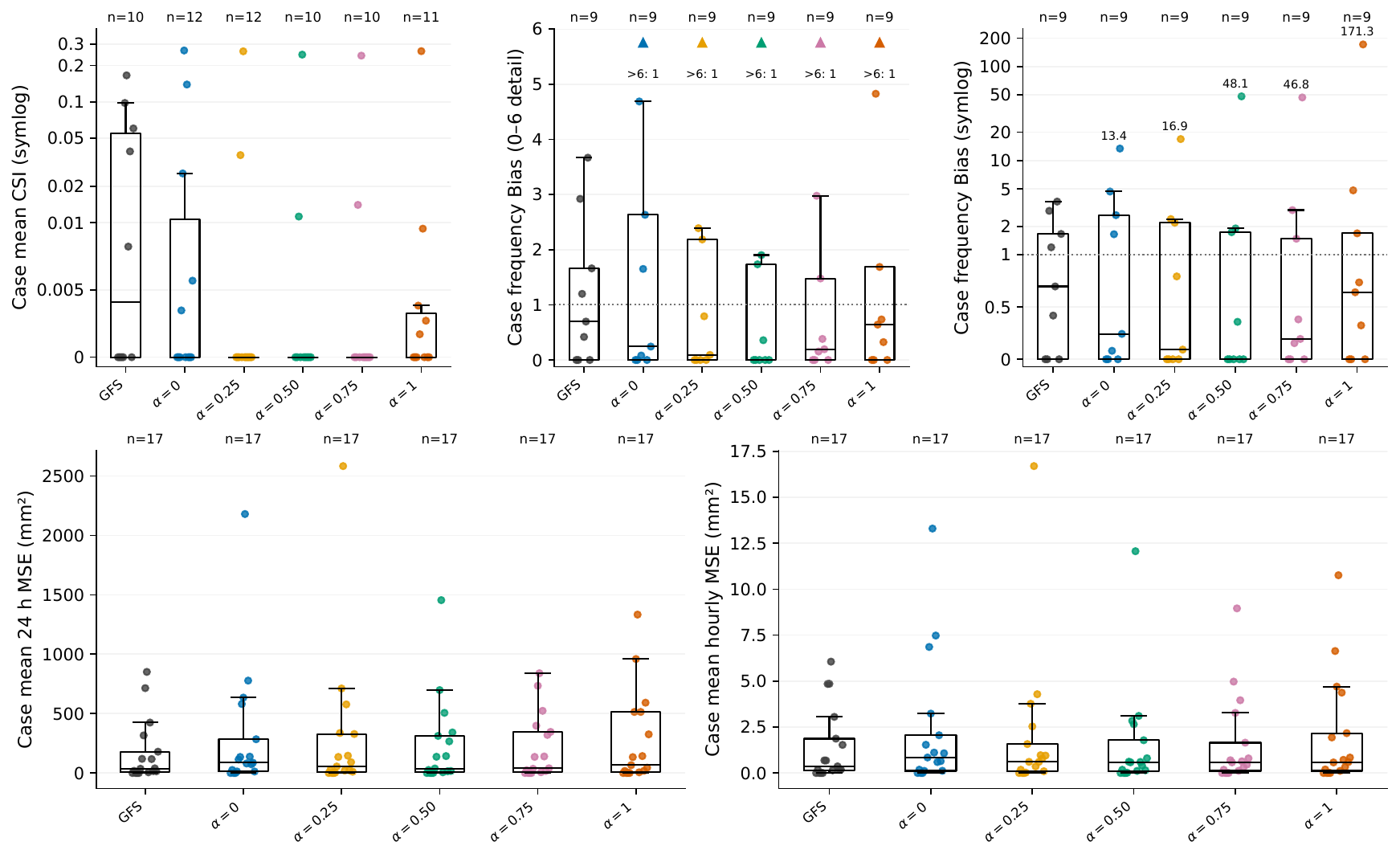}
\caption{Initialization-level distributions over the 17 starts. Boxes show quartiles and medians; whiskers extend to observations within 1.5 interquartile ranges. CSI and full-range Bias use symmetric-log axes (base 10, linear scale 1), linear from zero to 0.01 and 1, respectively, and logarithmic above these thresholds. Zeros and all extremes are retained; box statistics use untransformed values. Both Bias panels show the same nine defined cases per driver. The 0--6 detail remains linear; triangles mark values above its range. The dotted line is Bias=1. Counts denote valid cases, not independent events.}
\label{fig:boxes}\end{figure}

Table~\ref{tab:sd} reports descriptive initialization-level means and standard deviations. For example, 24-hour MSE across starts is $302.20\pm541.73$ mm$^2$ for unadapted Aurora, compared with $209.57\pm273.28$ mm$^2$ at strength 0.75. The distributions in Figure~\ref{fig:boxes} are strongly skewed, and the largest cases cannot be summarized adequately by an aggregate mean alone.

\begin{table}[t]\centering\small
\caption{Initialization-level mean $\pm$ sample SD; parentheses give valid initialization counts. CSI and Bias here are \emph{case statistics}, not the headline pooled estimators. MSE$_{24}$ averages the 49 windows; MSE$_1$ averages hours 1--72.}
\label{tab:sd}
\setlength{\tabcolsep}{3pt}
\resizebox{\linewidth}{!}{\begin{tabular}{lrrrr}\toprule
Driver & Case CSI & Case Bias & MSE$_{24}$ (mm$^2$) & MSE$_1$ (mm$^2$)\\\midrule
\tableinput{tables/dispersion.tex}
\bottomrule\end{tabular}}
\end{table}

Only nine starts contain reference heavy-rainfall events under the main-window definition, so case Bias has $n=9$. Case CSI can additionally be defined when no reference event occurs but the forecast produces a false alarm; it is then zero. With neither reference nor forecast events, CSI is undefined. Thus effective case counts differ across drivers. Excluding undefined cases from this supplementary distribution does not remove their false alarms from the primary pooled evaluation.

\subsection{Spatial examples and the extended range}
Appendix~\ref{app:maps} compares three representative initializations. Within each date, the displayed window maximizes reference area-weighted precipitation over starts 0--48; it is not selected by forecast skill. All six drivers share a color scale and verification support. These examples illustrate differences in rainfall placement, spatial extent, and intensity across the drivers; they are descriptive comparisons rather than a controlled attribution of physical mechanisms.

For window starts 0--72, mean CSI is lower for every driver (Appendix~\ref{app:extended}). This range adds starts whose rain ends at 73--96 hours. GFS has 70 valid CSI windows and each Aurora strength has 73. GFS's last three zero-denominator windows are undefined rather than missing forecasts. They are not filled. The effective averaging ranges differ and are disclosed explicitly. The longer-range results support the same broad trade-off but should not be represented as formal forecast skill confined to an ending lead of 72 hours.

\section{Discussion and limitations}
The experiment demonstrates a mismatch between upstream adaptation objectives and downstream precipitation criteria. The fitted state loss gives additional weight to environments associated with heavy rain, but neither predicts rainfall directly nor optimizes WRF rainfall skill. Changing atmospheric humidity, winds, pressure, or temperature can change transport and convective responses. The observed non-monotonic response to $\alpha$ is compatible with such a nonlinear coupling, but rainfall scores alone do not identify the responsible variable, boundary imbalance, or physical process.

Reduced whole-domain error and higher CSI answer different questions. Conservatively weak rainfall may incur smaller squared errors in dry or weakly raining areas while missing threshold exceedances. Conversely, a broader predicted rain footprint may raise detection and false-alarm counts together. GFS's lower MSE is therefore not inconsistent with its lower CSI. Likewise, strength 0.75 can improve magnitude/frequency behavior relative to the unadapted Aurora while losing event overlap. These findings motivate downstream-aware evaluation, not an unsupported assertion that one objective is universally preferable.

Several limitations delimit the result. First, ERA5 precipitation is the only validation reference used here. Its errors and effective resolution affect the apparent skill, and training/verification on the same reanalysis family limits independence even with distinct years. CMA, CPC, and CHIRPS validation is planned but has not been performed. Such validation must respect each product's spatial support, accumulation interval, and reporting time; daily products cannot establish hourly moving-window truth by temporal interpolation.

Second, skill varies across precipitation cases, and leave-one-initialization-out diagnostics show that the aggregate Aurora--GFS CSI ordering can reverse with sample composition. The reported ordering is therefore specific to this diagnostic sample; broader generalization requires verification on more independent events. Detailed case and sensitivity tables are retained in the reproducibility package.

Third, the GFS control shares retrospective initialization and a restricted interface with Aurora. It does not establish superiority over a fully configured operational NWP system. The auxiliary persistence and diagnostic pressure/humidity rules should be tested separately before deployment. Differences between the older and current convection implementations also make the exact WRF version important; no cross-version or alternative-physics causal claim is made.

Fourth, verification occurs on a coarse common grid after conservative averaging. The work does not demonstrate 1-km rainfall skill, gauge-scale extremes, ensemble reliability, or an end-to-end optimized forecasting system. Future improvements should prioritize independent precipitation references, predeclared calibration/test separation, and physical diagnostics of boundary-state changes before claiming a robust skill gain. Additional training sophistication is not itself evidence of better precipitation forecasts.

\section{Conclusion}
A single precipitation-weighted LoRA adaptation of Aurora changes WRF rainfall in a strength-dependent, metric-dependent way. Over 17 retrospective starts, unadapted Aurora has the largest pooled-window mean CSI, whereas an intermediate strength substantially reduces its continuous precipitation error and frequency overprediction. GFS remains best in whole-domain MSE under the common coupling rules. Within this controlled experiment, adaptation strength provides a means of adjusting the balance between detection, frequency calibration, and intensity error rather than improving all criteria simultaneously.

\section*{Data and code availability}
Ancillary files provide verification-grid precipitation arrays, analysis and plotting code, numeric tables, bootstrap statistics, archive hashes, and reproduction instructions. They reproduce the reported analysis, not model training or forecast integration; trained weights, full training data, and native WRF histories are not included. ERA5 and GFS data are available from their providers \citep{era5,gfs}, and Aurora code and pretrained checkpoints from Microsoft \citep{aurora}. ERA5-derived arrays are credited to the Copernicus Climate Change Service and ECMWF; source and licence details accompany the ancillary data.

\clearpage
\bibliographystyle{plainnat}
\bibliography{references}

\clearpage
\appendix
\section{Evaluation sample}\label{app:cases}
\begin{table}[h]\centering\small
\caption{The 17 evaluation initializations (00 UTC, 2023). Reference statistics use the fixed 356-cell mask and 49 rolling 24-hour windows starting at 0--48 h. Event counts use the 50-mm threshold and include overlapping windows; zero denotes no threshold exceedance, not necessarily no rain.}
\begin{tabular}{lrrr}\toprule
Date & Event cell-windows & Max event cells/window & Max 24 h (mm)\\\midrule
\tableinput{tables/cases.tex}
\bottomrule\end{tabular}\end{table}

\medskip
\section{Extended window-start range}\label{app:extended}
\begin{table}[!ht]\centering\small
\caption{Supplementary starts 0--72 h, requiring rain forecasts through 96 h. $n$ counts defined pooled CSI windows out of 73. Bias retains all false alarms, including in windows without reference events.}
\begin{tabular}{lrrrr}\toprule
Driver & Mean CSI & $n$ & Pooled Bias & 24 h MSE (mm$^2$)\\\midrule
\tableinput{tables/extended.tex}
\bottomrule\end{tabular}\end{table}

\clearpage
\section{Representative precipitation fields}\label{app:maps}
\begin{figure}[!ht]\centering
\includegraphics[width=\linewidth]{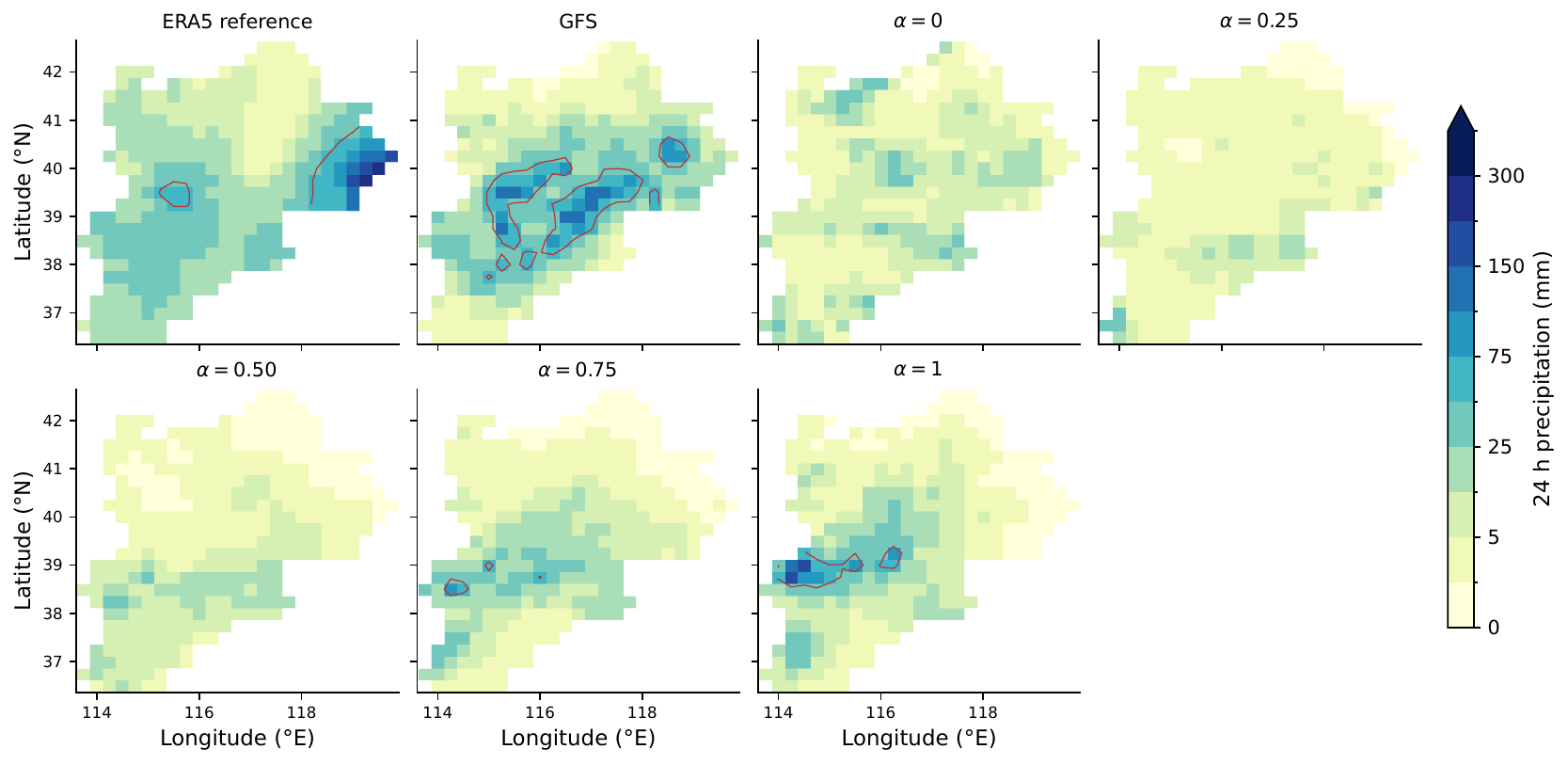}
\caption{20 July 2023 initialization, accumulation from +25 to +49 h. The window is chosen from reference precipitation only. Fields are conservatively mapped to the verification grid; red contours mark 50 mm. The same discrete color scale is used for all drivers and all three map figures.}
\label{fig:map1}\end{figure}
\begin{figure}[!ht]\centering
\includegraphics[width=\linewidth]{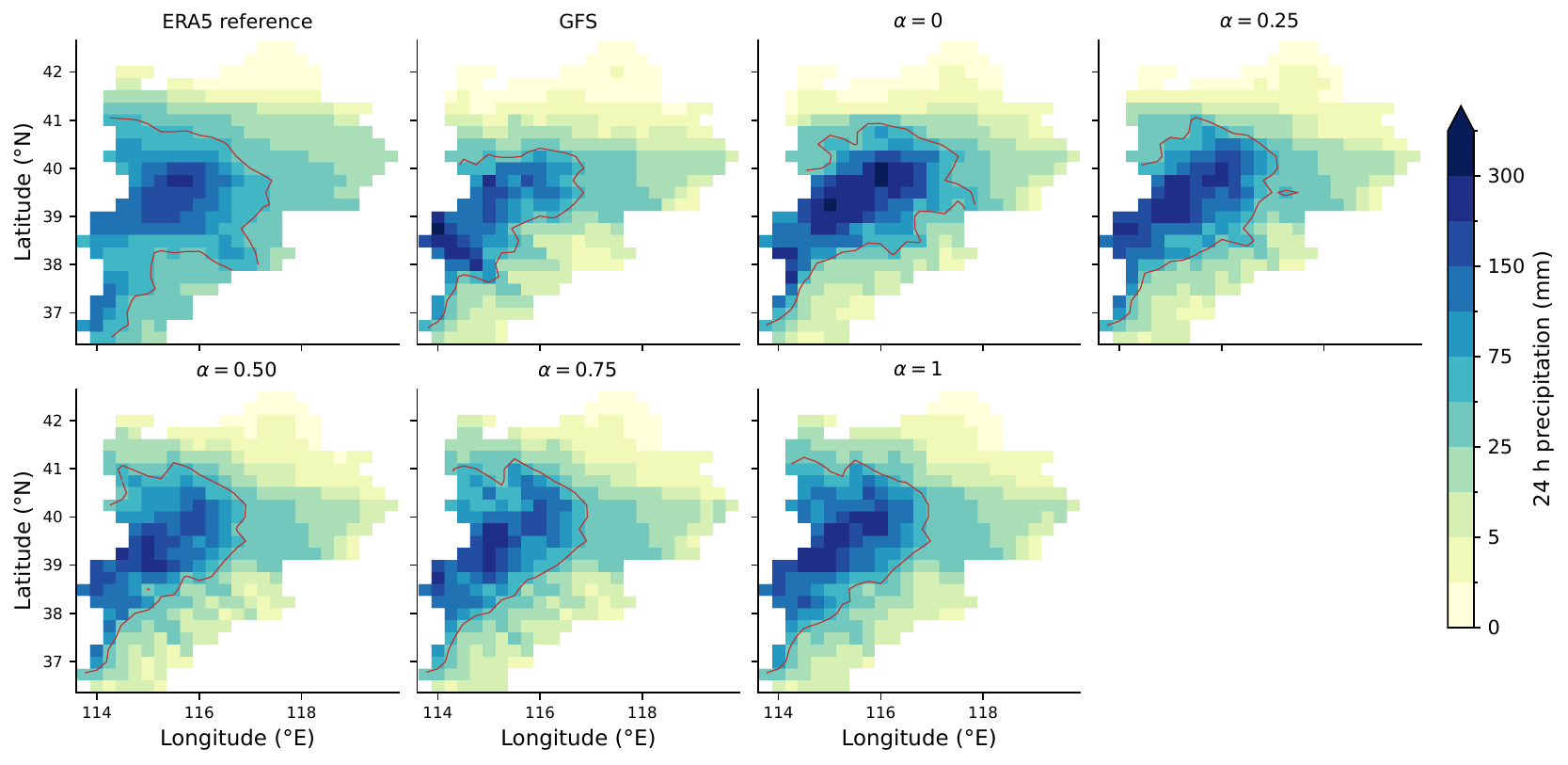}
\caption{30 July 2023 initialization, accumulation from +0 to +24 h, selected by the same reference-only rule. Fields use the common verification grid, color scale, and 50-mm contour convention.}
\label{fig:map2}\end{figure}
\begin{figure}[!ht]\centering
\includegraphics[width=\linewidth]{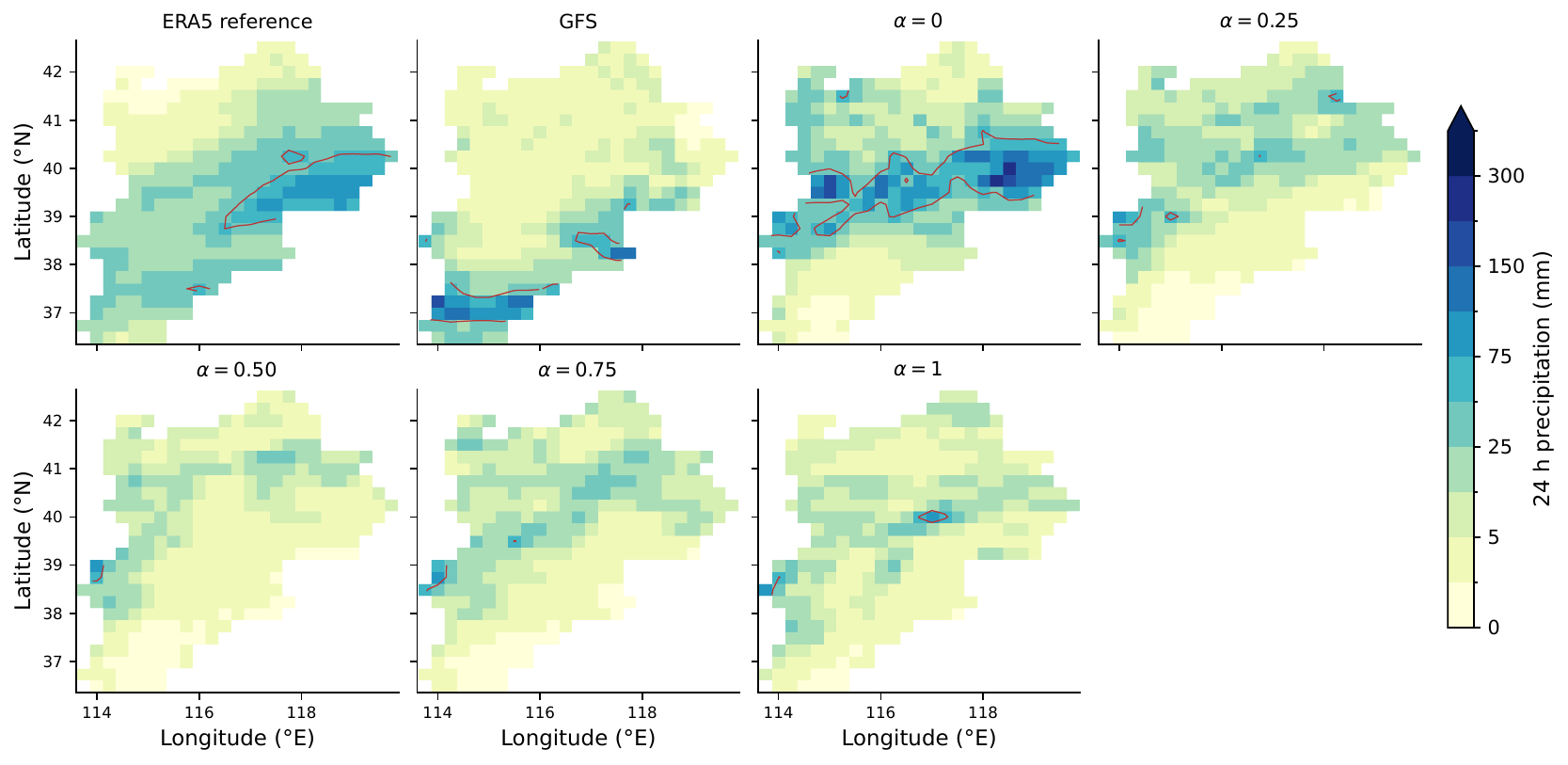}
\caption{10 August 2023 initialization, accumulation from +30 to +54 h, selected by the same reference-only rule. All six drivers are displayed on the common verification grid and color scale.}
\label{fig:map3}\end{figure}
\end{document}